\documentclass[pra,twocolumn,epsfig,rotate,superscriptaddress,showpacs,amsmath,amsfonts,amssymb,floatfix]{revtex4}
\usepackage{graphicx}
\usepackage{dcolumn}
\usepackage{bm}
\newcommand{\tp}{\hat{\tilde{p}}}
\newcommand{\tk}{\tilde{K}}

\newcommand{\ZM}{\mathbb Z}
\newcommand{\tilt}{{\hat H}_{\text{\tiny tilt}}}

\begin{document}

\title{Long-lasting Exponential Spreading in Periodically Driven Quantum Systems}
\author{Jiao Wang}
\affiliation{Department of Physics and Institute of Theoretical Physics and Astrophysics,\\
Xiamen University, Xiamen 361005, China}
\author{Italo Guarneri}
\affiliation{ Center for Nonlinear and Complex Systems,
Universit\`a degli Studi dell'Insubria, Via Valleggio 11, 22100 Como, Italy}
\affiliation{Istituto Nazionale di Fisica Nucleare, Sezione di Pavia,
via Bassi 6, 27100 Pavia, Italy}
\author{Giulio Casati}
\affiliation{ Center for Nonlinear and Complex Systems,
Universit\`a degli Studi dell'Insubria, Via Valleggio 11, 22100 Como, Italy}
\affiliation{Consorzio Nazionale Interuniversitario per le Scienze Fisiche della Materia, e CNR-INFM}
\affiliation{Istituto Nazionale di Fisica Nucleare, Sezione di Milano,
via Celoria 16, 20133 Milano, Italy}
\author{Jiangbin Gong}
\affiliation{Department of Physics and Center for Computational Science and Engineering, National University of Singapore, Singapore 117542, Republic of Singapore}
\affiliation{NUS Graduate School for Integrative Sciences and
Engineering, Singapore 117597, Republic of Singapore}
\date{\today}
\begin{abstract}
Using a dynamical model
 relevant to cold-atom experiments, we show that long-lasting exponential spreading of wave packets in momentum space is possible.  Numerical results are explained via a pseudo-classical map, both qualitatively and quantitatively. Possible applications of our findings are also briefly discussed.
\end{abstract}
\maketitle

Apart from idealized exceptions like inverted oscillators \cite{inverted}, exponential growth of observable quantities is thought to be
a rare occurrence in small quantum systems. Exponential instability, which is  the landmark of classical chaos, does not give rise to long lasting, physically observable exponential growth, not even in classical mechanics, because exponentially fast motion along unstable manifolds is concealed as such manifolds fold in homoclinic or heteroclinic tangles. Moreover, it does not survive in quantum mechanics beyond a short time scale $t_E\approx -\ln(\hbar)/\lambda$~\cite{casatibook}, where  $\lambda$ is the maximal Lyapunov exponent.
Using a simple dynamical model relevant to today's cold-atom experiments, we show in this work that exponential quantum spreading (EQS) of wave packets in momentum space does exist for a significantly long time scale. This is achieved by exploiting two intriguing mechanisms at once.  First,  our quantum system is set close to a quantum resonance condition, in a ``pseudo-classical" (yet strictly quantum) regime where the quantum dynamics follows trajectories of a \emph{fictitious} classical system, which emerges at small values of  a ``pseudo-Planck" constant $\hbar^*$. This ``constant" $\hbar^*$ measures the detuning from exact quantum resonance and
can easily be made quite small by varying physical parameters,
while keeping the system in a deeply quantum regime.
Second, appropriate driving fields are chosen such that  a pseudo-classical stable manifold of an unstable fixed point
can support easy-to-prepare quantum states.
As shown below, these factors concur in  eventually generating EQS in momentum space (or exponential energy increase) as a  transient but long-lasting
dynamical phenomenon,
whose duration is given by $(A/{\hbar^*})\ln(C/{\hbar^*})$ ($A$ and $C$ are constants for small $\hbar^*$) and therefore can be enlarged   to a very long time scale as parameter $\hbar^*$ decreases. The exponential rate of EQS scales linearly with $\hbar^*$ and is typically small for close-to-resonance situations, yet the EQS momentum scale can be huge because it scales with $1/\hbar^*$.
Such long-lasting EQS
does not rest on the chaotic motion which is present in the {\it classical} limit, but  on quasi-integrable motion which is instead present in the {\it pseudo-classical} limit.

The dynamical model used here is a simple modification of the well-known kicked-rotor model~\cite{casatibook}, a paradigm of quantum chaos that has been extensively studied both theoretically and experimentally.  For all previous variants of such type of driven systems, quantum spreading  is known to range from ballistic propagation to super-diffusion as well as standard linear diffusion followed by dynamical localization~\cite{casatibook}.  Our finding of long-lasting EQS is thus an intriguing result in quantum chaos.  In a broader context, the EQS reported here
is different from other forms of fast wave-packet spreading like, e.g., the ``super-ballistic" diffusion found in a lattice model~\cite{ketz} (where the wavepacket spreading can be a cubic function of time) and the transient ``hyperdiffusion" found in generalized Brownian motion~\cite{hanggi} (where a square variance may increase at the fifth power of time).  Furthermore, analogous to the novel exponential acceleration of classical particles~\cite{shah}, the possibility of long-lasting EQS might be useful for designing some robust acceleration methods for quantum systems.

Consider then the following Floquet propagator for a double-kicked-rotor model
\begin{equation}
\hat{U}=e^{-i(T-T_0)\frac{\hat{p}^2}{2\hbar}}e^{-i\frac{K}{\hbar}\cos(\hat{q})}
e^{-iT_0\frac{\hat{p}^2}{2\hbar}}e^{-i\frac{K}{\hbar}\cos(\hat{q})},
\label{DKR}
\end{equation}
with all quantities in dimensionless units. Throughout $\hat{q}$ and $\hat{p}$ are the canonical  coordinate and momentum operators.  This system was first experimentally studied in Ref.~\cite{monteiro} by subjecting cold atoms to two period-$T$ $\delta$-kicking sequences of an optical lattice potential, time-shifted by $T_0<T$ with respect to each other.  Under the so-called quantum main resonance condition $T\hbar=4\pi$, achievable in a number of cold-atom experiments~\cite{raizen,resonanceexp,steinberg,newzealand,summy}, we obtain an ``on-resonance double-kicked rotor" model (RDKR) ~\cite{gong-wang-pre2007,wang-gong-pra2008}, described by the Floquet propagator:
\begin{equation}
\hat{U}_{\text{\tiny R}}=e^{-i (\frac{4\pi}{\hbar}-T_0)\frac{\hat{p}^2}{2\hbar}}e^{-i\frac{K}{\hbar}\cos(\hat{q})}
e^{-iT_0\frac{\hat{p}^2}{2\hbar}}e^{-i\frac{K}{\hbar}\cos(\hat{q})}.
\label{ondr}
\end{equation}
Due to spatial periodicity of the kicking potential $\cos(q)$, changes in momentum $p$ must occur in integer multiples of $\hbar$. The momentum eigenvalues of the system are hence given by $m\hbar+\beta\hbar$, where $\beta\in [0,1)$ is a quasi-momentum variable determined by the initial state and conserved by the dynamics. In cold-atom experimental realizations  of the kicked-rotor model, the initial state is a mixture of states with a certain spread of $\beta$, where the spread can be made small by loading a Bose-Einstein condensate into the lattice such that the initial-state coherence spans over many lattice constants~\cite{resonanceexp,newzealand,summy}. For $\beta=0$, the Floquet spectrum of $\hat{U}_{\text{\tiny R}}$ plotted against $T_0\hbar\in [0,4\pi]$ yields the famous Hofstadter's butterfly structure~\cite{wang-gong-pra2008}, with the spectra identical with that of the kicked-Harper model~\cite{harper}
for arbitrary irrational values of $T_{0}\hbar/\pi$~\cite{JMP}.

\begin{figure}
\vskip-0.3cm
\hspace{-0.5cm}\includegraphics[width=8.4cm]{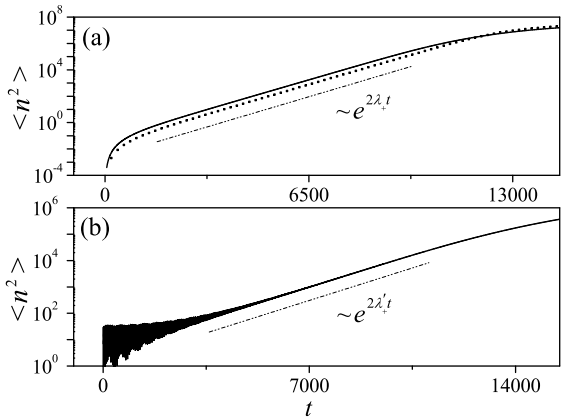}
\vskip-0.5cm
\caption{Two examples of long-lasting exponential quantum spreading (solid lines) in RDKR [see Eq.~(\ref{ondr})]. (a) $K=5$,
$T_0=1$, and $T_0\hbar=2\pi+\hbar^*$ with $\hbar^*=10^{-3}$. Dotted line represents the pseudo-classical result and dash-dotted line represents $\sim e^{2\lambda_{+}t}$ with $\lambda_{+}$ found analytically. (b) $K=9$, $T_0=1$, and  $T_0\hbar=30\pi/13+1.7\times 10^{-4}$. Dash-dotted line represents the best exponential fitting (shifted) that suggests an exponential rate of  $\lambda_{+}'=4.1\times 10^{-4}$. Here and in all other figures, plotted quantities are in dimensionless units.}
\label{examples}
\end{figure}

Representative numerical EQS results for the quantum map $\hat{U}_{\text{R}}$ are depicted in Fig. 1, with the time dependence of the mean value $\langle n^2 \rangle\equiv \langle \hat{p}^2/\hbar^2 \rangle$ plotted in a log scale for a zero-momentum initial state
(hence $\beta=0$). As seen clearly from Fig.~1, apart from a short transient initial stage, $\log(\langle n^2 \rangle)$ displays a striking linear dependence on the number of iterations of $\hat{U}_{\text{\tiny R}}$ (denoted by $t$). That is, Fig.~1 demonstrates {\it exponential} spreading in  momentum space, with the exponential behavior sustaining over a wide range of time and momentum scales. For example, in Fig. 1(a), the exponential spreading occurs from $ t\approx 10^3$ up to $t\approx 10^4$, $\langle n^2\rangle $ increases from $10^{-1}$ to $10^6$.  For the case in Fig.~1(b), the exponential spreading also brings $\langle n^2\rangle $ from $10^2$ to $10^5$, over a time scale of the order of $10^4$. In all cases when such exponential behavior is observed in our extensive numerical studies, $\hbar T_0$ is close to $2\pi M/N$, with $M$ and $N$  being  odd integers. For instance  $\hbar T_0$ is close to $2\pi$ for the case in Fig.~1(a), whereas for the case of Fig.~1(b) it is slightly away from $30\pi/13$.

 To explain the EQS discovered in our numerical work, in our theoretical analysis we exploit
 the above-mentioned close-to-resonance condition. To that end we consider  cases when $\hbar T_0$ is close to a low-order resonance, i.e.,  $\hbar T_0=2\pi+\hbar^*$ with $|\hbar^*|\ll 1$. The reason of the notation $\hbar^*$ will become manifest in Eq.~(\ref{DKR1}). We note that under the precise condition $\hbar T_0=2\pi$ and $\beta=0$ we have a so-called quantum antiresonance, and that interesting aspects of long-time  quantum diffusion in the vicinity of quantum antiresonance were previously analyzed for a different model, describing a kicked charge in a magnetic field \cite{Danapre}.
The long-time asymptotic behavior of quantum motion is not, however, in the scope of the present work, as we are instead interested in the transient, though long-lasting EQS.  We now proceed by adopting a ``pseudo-classical" approximation which was successfully implemented in the study of other close-to-resonance phenomena observed in cold-atom optics~\cite{classical}. That is, quantum dynamics under close-to-resonance conditions, and hence far away from its proper classical limit, may be nevertheless extremely close to the behavior of a  fictitious classical system, unrelated to the classical limit.
 Cases with $\beta=0$ will be studied first.  For convenience we also set $\hbar^*>0$.  With such specifications, the first factor in  $\hat{U}_{\text{\tiny R}}$ in Eq.~(\ref{ondr}) reduces to $e^{iT_0\frac{\hat{p}^2}{2\hbar}}$,  a time-reversal of the normal free-rotor propagator and a feat experimentally achieved in Ref.~\cite{newzealand}. Next we introduce the rescaled kicking strength  $\tilde{K}\equiv K{\hbar^*}/{\hbar}$ and the rescaled momentum $\hat{\tilde{p}}\equiv \hat{p}{\hbar^*}/\hbar$. As $\beta=0$, $\tilde p$ is quantized in integer multiples of $\hbar^*$, so the relation $
e^{\pm iT_0\frac{\hat{p}^{2}}{2\hbar}} =
e^{\pm \frac{i}{\hbar^*}\left(\frac{\hat{\tilde{p}}^2}{2}+\pi \hat{\tilde{p}}\right)}
$ holds true, and using it in Eq.~(\ref{ondr}) yields
\begin{eqnarray}
\hat{U}_{\text{\tiny R}}^{(\beta=0)} & = & e^{\frac{i}{\hbar^*}\left(\frac{\hat{\tilde{p}}^2}{2}+\pi \hat{\tilde{p}}\right)}
e^{-\frac{i}{\hbar^*}\tilde{K}\cos(\hat{q})} \nonumber \\ & & \times  e^{- \frac{i}{\hbar^*}\left(\frac{\hat{\tilde{p}}^2}{2}+\pi \hat{\tilde{p}}\right)}
e^{-\frac{i}{\hbar^*}\tilde{K}\cos(\hat{q})}\;.
\label{DKR1}
\end{eqnarray}
 In Eq.~(\ref{DKR1}) ${\hbar}^*$ manifestly plays a Planck-constant-like role, so that in the ``pseudo-classical limit" when $\hbar^*\to 0$, the quantum dynamics described by (\ref{DKR1}) comes  closer and closer to that of a classical double-kicked rotor whose kinetic energy term alternates between $\pm(\frac{{\tilde{p}}^2}{2}+\pi {\tilde{p}})$ but with a fixed kicking potential $\tilde{K}\cos(q)$.  The associated pseudo-classical map can then be obtained directly:
\begin{eqnarray}
{q}_{n+1} &= & {q}_{n}+\tilde{K} \sin(\tilde{p}_{n+1}+q_{n+1}) \nonumber \\
(\tilde{p}_{n+1}+q_{n+1})&= &(\tilde{p}_{n}+q_{n})+ \tilde{K}\sin(q_n).
\label{cmap}
\end{eqnarray}
Remarkably, this map can be recognized as a kicked-Harper model \cite{harper}, with the canonical pair identified as
$q$ and $\tilde{p}+q$. This map has  unstable fixed points at $q=m\pi$, and $\tilde{p}+q=n\pi$, with $m+n$ being an even number, and Lyapunov exponents $\lambda_{\pm}= \ln\left[\left(\tilde{K}^2+2\pm\sqrt{\tilde{K}^4+4\tilde{K}^2}\right)/2\right]$.  In the limit of $\tilde{K}\rightarrow 0$, stable and unstable manifolds of such fixed points connect in a periodic network, and phase-space is divided into cells in the shape of parallelograms, with two sides parallel to the $q$ (zero-momentum) axis, and the other two inclined by $\arctan(2)$. For small nonzero $\tilde K$, splitting of such separatrices occurs, giving birth to an exponentially thin stochastic web.
Figure 2(a) depicts a  typical phase space structure, which is indeed a portrait deformed from that of a kicked-Harper model~\cite{harper}. The almost straight lines in Fig.~2(a) are the stochastic web.

\begin{figure}
\vskip-0.cm
\hspace{-0.55cm}\includegraphics[width=8.4cm]{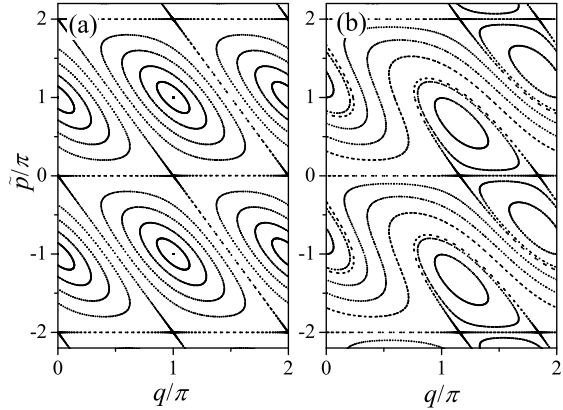}
\vskip-0.5cm
\caption{
Phase space portrait of the pseudo-classical limit of RDKR for $K=5$, $T_0=1$,  and $T_0 \hbar=2\pi+\hbar^*$ with $\hbar^*=3\sqrt{2}\times 10^{-4}\pi$. $\beta=0$ in (a) and  $\beta=4\times 10^{-5}$ in (b).}
\label{portrait}
\end{figure}

The exponential spreading seen in Fig.~1(a) can now be understood with our pseudo-classical map. The initial state $p=0$ corresponds to an ensemble placed almost perfectly astride  the stable manifold of the unstable fixed point at $(q,\tilde{p})=(\pi,0)$.
The dynamics first pulls this ensemble towards this fixed point, and then repels the ensemble away from the fixed point at the exponential rate $\lambda_{+}$ along the unstable manifold. A key point is that, although this exponential increase involves a maximum change of $2\pi$ of the pseudo-classical momentum $\tilde p$, it  translates into exponential behavior  of the physical momentum  $p$ over a huge  scale for $|\hbar^*|\ll1$, because $p\sim \tilde{p}\hbar/\hbar^*$.

Figure~1(a) shows that results of pseudo-classical and quantum calculations nicely agree on the ensemble level.
It is interesting to note that because the separatrix structure in the standard kicked-Harper model~\cite{harper} is tilted away (45 degrees for vanishing kicking strength) from the zero-momentum axis, our exponential spreading mechanism cannot be generated there, because a state sitting on the stable branch of a separatrix  would  be delocalized on a huge momentum scale right at the start.
It should also be stressed that this mechanism does not rest on chaos; although chaotic diffusion inside the thin stochastic web is itself an intriguing  phenomenon~\cite{dana}, EQS  is instead an outcome  of quasi-integrability.

The time scale of the found exponential spreading, denoted by $t_{\text{exp}}$, can also be estimated. Here we sketch a crude heuristic argument, leaving a more formal version for the supplementary material~\cite{supp}. For a pseudo-classical ensemble aligned with the $q$ (zero-momentum) axis with a momentum width $\delta\tilde{p}$, exponential momentum growth $\tilde{p}(t) \propto \delta \tilde{p}\exp(\lambda_{+} t)$ along the unstable manifold  continues until $\tilde{p}(t)\sim \pi$ , and therefore
$t_{\text{exp}}\sim\lambda_{+}^{-1}\ln(\pi/\delta\tilde{ p})$. It can be shown that $\delta\tilde{p}\propto\hbar^*$ for the zero momentum quantum initial state, so, on account of $\lambda_+\approx \tilde{K}=K\hbar^*/\hbar$ (for small $\tilde{K}$), we obtain $t_{\text{exp}} \sim (\hbar/K\hbar^*)\ln(\pi/\hbar^*)$. Figure 3 shows that this theoretical scaling with $\hbar^{*}$ is in excellent agreement with the numerically found time scale of EQS.  Additional numerical results (not shown) of the $K$-dependence of $t_{\text{exp}}$ also agree with our theory. It is now clear
 that as $\hbar^*$ decreases and exact antiresonance is approached,  $t_{\text{exp}}$ increases and $\lambda_{+}$ decreases~\cite{danapre1}.

\begin{figure}
\vskip0.2cm
\includegraphics[width=8cm]{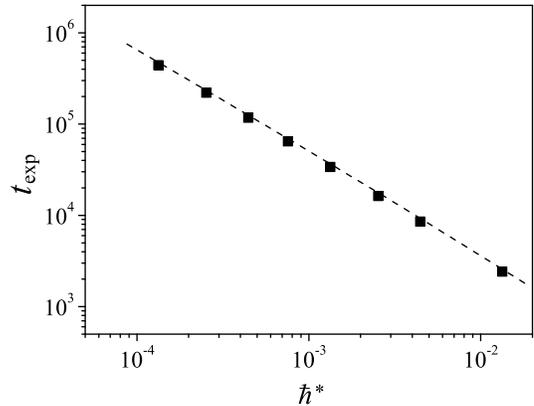}
\vskip-0.5cm
\caption{Numerical $t_\text{exp}$ (after which the spreading in momentum space deviates from the exponential law) for
RDKR (black squares) with $K=1$,
$\beta=0$, $T_0 \hbar=2\pi+\hbar^*$ and a varying $\hbar^*$. The dashed line represents the theoretical prediction.} 
\label{ratchet}
\end{figure}

To motivate experimental interest we now examine the effects of a non-zero quasi-momentum $\beta$.
Let us generalize our previous definition of $\tilde{p}$, i.e., $\tilde{p}\equiv |\hbar^*|(p-\beta\hbar)/\hbar.$  Still focusing on $T_0 \hbar=2\pi+\hbar^*$ as an example, we find the following pseudo-classical map associated with $U$ at $\beta\neq 0$:
\begin{eqnarray}
\tilde{p}'_n&=&\tilde{p}_n + \tilde{K}\sin(q_n),\nonumber \\
q'_n& =& q_n + \tilde{p}'_n+(2\beta+1)\pi,\nonumber \\
\tilde{p}_{n+1}&=&\tilde{p}'_n + \tilde{K}\sin(q'_n), \nonumber \\
q_{n+1}&=&q'_n- \tilde{p}_{n+1}+(2\beta+1)\pi.
\label{betamap}
\end{eqnarray}
Here $\tilde{p}_n'$ and $q_n'$ are two intermediate variables. 
 Note also that this mapping reduces to Eq.~(\ref{cmap}) if $\beta=0$. The phase space structure of the mapping in Eq.~(\ref{betamap}) is still periodic in ${\tilde{p}}$ with a period $2\pi$, with one example shown in Fig.~2(b). Within the regime $-\pi\le \tilde{p}<\pi$, there are two unstable fixed points located at $\tilde{p}=2\beta\pi$, $q=2\pi -\xi$ or $q=\pi+\xi$, with $\xi\equiv \arcsin(4\pi\beta/\tilde{K})$.  As indicated by Fig.~2(b), for a sufficiently small $\beta$, the stable direction of the two unstable fixed points is still almost parallel to the $q$ axis and the unstable direction is still extended in $\tilde{p}$. The exponential spreading mechanism hence survives. The exponential divergence rate can  be found from a linear stability analysis of the fixed points. A simple calculation
gives that the rate is now dependent on $\beta$, i.e., $\lambda_{+}(\beta)= \ln\left[\left(\tilde{K'}^2+2+\sqrt{\tilde{K'}^4+4\tilde{K'}^2}\right)/2\right]$, with $\tilde{K'}^2=\tilde{K}^2-16\pi^2\beta^2$.  As such, $\lambda_{+}(\beta)$ is maximal for $\beta=0$. This indicates that in the  presence of a distribution of values of $\beta$ centered at $\beta=0$,
the spreading dynamics will be dominated by
the $\beta=0$ component.
On the other hand, when $|\beta|\ge \beta_{\text{cr}}\equiv \tilde{K}/4\pi=K\hbar^*/(4\pi\hbar)$  the fixed points disappear, and then the pseudo-classical mechanism for EQS is lost.  

In Fig.~4(a) we numerically examine how a nonzero $\beta$ affects the exponential spreading behavior. There it is seen that an exponential spreading law eventually turns into a periodic oscillation as $\beta$ increases.  Using this transition we have also numerically extracted critical $\beta$ values as a function of $K$ and then compared them with our theoretical analysis of $\beta_{\text{cr}}$.  Such a comparison is shown in Fig.~4(b), with excellent agreement. The important message from Fig.~4 is then the following: to observe exponential spreading with nonzero $\beta$ values, it is advantageous to increase $\beta_{\text{cr}}$, which can be achieved by increasing the kicking strength $K$.  Assuming that experimental uncertainties in $\beta$ is around $\Delta\beta =10^{-3}$  and that the near-resonance condition can be achieved at an accuracy of $\hbar^*/(T_0 \hbar)\approx 10^{-3}$,  then this requires $KT_0\ge 4\pi$, which is in the appropriate regime for current cold-atom experiments~\cite{resonanceexp,steinberg,newzealand,summy}.

\begin{figure}
\vskip-0.2cm
\hspace{-0.5cm}\includegraphics[width=8.6cm]{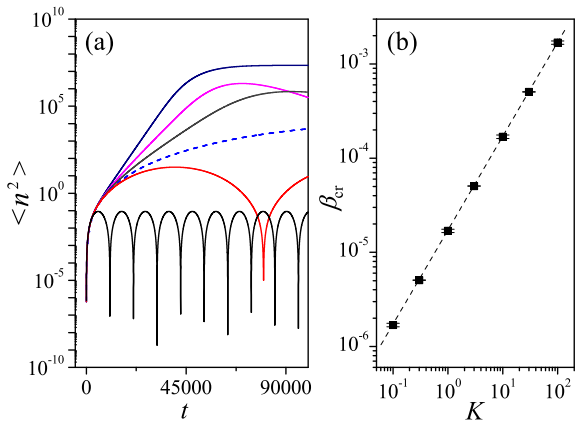}
\vskip-0.5cm
\caption{(Color online)(a) Effects of nonzero $\beta$ on exponential quantum spreading. $K=1$, $T_0=1$, and $T_0 \hbar=2\pi+\hbar^*$ with $\hbar^*=3\sqrt{2}\times 10^{-4}\pi$. Solid lines (from top to bottom) are for $\beta/10^{-5}=0$, $1.2$, $1.5$, $1.8$ and $5$, respectively. Dashed line is for $\beta=\beta_{\text{cr}}\approx 1.688\times 10^{-5}$.  The initial state is a momentum eigenstate with $p=\beta \hbar$. (b) Comparison of the numerical $\beta_{\text{cr}}$ (black squares) with theoretical $\beta_{\text{cr}}$ for $\hbar^*=3\sqrt{2}\times 10^{-4}\pi$ and a varying $K$.}
\label{n2tbeta}
\end{figure}

As illustrated in Fig.~1(b), EQS is also observed near higher-order resonances, i.e., $T_0\hbar\simeq 2\pi M/N$.
There the pseudo-classical approach requires nontrivial modifications~\cite{hires}, which are not examined in the present work.
Finally, we briefly mention some interesting applications of the EQS found here. First,
if we introduce a phase shift between the two $\delta$-kicking potentials as in Ref.~\cite{ratchet} and then tune $T_0\hbar$ close to a resonance
condition, we find that not only $ \langle n^2\rangle $ but even the current (i.e. the momentum expectation value) can  increase exponentially over a long time scale.
A novel type of ratchet accelerator can hence be formed.  Second, because the dynamical model used here can be equally realized in spin chain models kicked by an external parabolic magnetic field~\cite{bose,gong-wang-pre2007}, an exponential propagation of quantum excitation along a spin chain is in principle possible.

JW acknowledges the support by the NNSF (Grant No. 10975115) and
SRFDP (Grant No. 20100121110021) of China.  GC and IG acknowledge the support by the MIUR-PRIN 2008 and by Regione Lombardia. JG was supported by ARF Tier I, MOE of Singapore (grant No. R-144-000-276-112).

\vspace{1cm}
\appendix{ \begin{large}\textbf{Supplementary Material}\end{large}}

\vspace{0.5cm}

In this supplementary material we elaborate on the exponential quantum spreading (EQS) time scale, defined as $t_{\text{exp}}$ in the main text.
The initial ensemble $p=0$ is almost perfectly aligned with the stable manifold of the unstable fixed point at $(q,\tilde{p})=(\pi,0)$.
The ensuing dynamics is produced by the motion along invariant tori close to the separatrix [see Fig.~2(a)]: the dynamics
initially attracts the ensemble to the fixed point and then repels the ensemble away from the fixed point at an exponential rate.
Indeed, these tori close to the separatrix can support Floquet eigenstates, and the initial quantum state overlaps many of them.

Our estimate of $t_{\text{exp}}$ consists of three steps. First of all, quantized tori are identified with eigenstates of a so-called ``tilted Harper Hamiltonian" (tHH) (see later). This is correct at the lowest perturbative order.  Let $|0\rangle$ denote the state of (angular) momentum $0$, {\it i.e.}, the initial quantum state. It overlaps the eigenstates of tHH whose energies lie in a window of $(-\delta E,0)$. An estimate for $\delta E$ is derived by computing the spread of tHH in the state $|0\rangle$. As detailed below,  $\delta E$ is given by:
\begin{gather}
     \delta E\;\equiv\;\sqrt{\langle 0|\tilt^2|0\rangle\;-\;\langle 0|\tilt|0\rangle^2}=\;\frac12\tk\hbar^*\;.
     \label{dele}
     \end{gather}
     Second, we estimate the classical half-period $T(E)$ of motion along tori in the vicinity of the separatrix. To this end we  shall use the Kicked-Harper Hamiltonian in continuous time. This is correct at the lowest perturbative order and the separatrix is at energy $E=0$. We then obtain $T(E)$ for $E<0$ and $|E|\ll 1$.  It will be shown that $T(E)$  diverges logarithmically as the separatrix is approached, and at the leading order
\begin{equation}
\label{tde}
T(E)\;\sim\;\tk^{-1}\ln(-2{\tk}/{E})\;.
\end{equation}
Finally we identify $t_{\text{\tiny exp}}$ with $T(c\delta E)$,
where $c$ is an adjustable fitting parameter. Hence we arrive at the following scaling discussed in our manuscript:
\begin{gather}
  t_{\text{\tiny exp}}\;\sim \;\tk^{-1}\ln(\frac{A}{\hbar^{*}})\sim \frac{\hbar}{K\hbar^{*}} \ln(\frac{A}{\hbar^{*}})\;
\end{gather}
with $A=4/c$. Next we sketch the calculations leading to Eq.~(\ref{dele}) and (\ref{tde}).

{\it Energy width of initial state} -- The quantum double kicked rotor map $\hat{U}^{(\beta=0)}_{\text{R}}\;=\;\hat{V}^{-1}\hat{U}_0\;\hat{V}\;\hat{U}_0$ with
\begin{gather}
\hat{U}_0\;=\;e^{-\frac{i}{\hbar^*}\tk\cos(\hat{q})}\;,\;\;
\hat{V}\;=\;e^{-\frac{i}{\hbar^*}(\tp^2/2+\pi\tp)}
\end{gather}
may also be written as
$$
\hat{U}^{(\beta=0)}_{\text{R}}\;=\;e^{-\frac i{\hbar^*}\hat{H_1}}\;e^{-\frac i{\hbar^*}\hat{H_0}}
$$
where
\begin{eqnarray}
\hat{H_0}&=&\tk\cos(\hat{q}), \nonumber \\
\hat{H_1} &=& \hat{V}^{-1}\;\hat{H_2}\;\hat{V}.
\label{htwo}
\end{eqnarray}
A Floquet Hamiltonian can be defined as a self-adjoint $\hat {H}_f$ such that $\exp(-{i\hat{H}_f}/\hbar^*)=\hat{U}^{(\beta=0)}_{\text{R}}$. None is explicitly known about $\hat {H}_f$. However for small kicking strength an approximation is provided (via Trotter-Kato) by
$$
\hat {H}_f\approx \tilt\;=\;\hat{H}_0\;+\;\hat{H}_1,
$$
which may be termed the ``tilted Harper Hamiltonian" (tHH). Its pseudo-classical limit is obviously
\begin{equation}
\label{clthh}
H_{\text{\tiny tilt}}\;=\;\tk\bigl[\;-\cos(\tilde{p}+q)\;+\;\cos(q)\;\bigr]\;,
\end{equation}
however it does not really coincide with the back-quantization of it. Its matrix elements in the eigenbasis $|n\rangle$ $(n\in\ZM)$ of the angular momentum $\tp$ are easily computed noting that
$$
\hat{H}_0\;=\;\frac12\tk\;\sum\limits_n\bigl(\;|n\rangle\langle n+1|\;+\;|n\rangle\langle n-1|\;\bigr)\;,
$$
whence, using Eq.~(\ref{htwo}),
\begin{eqnarray}
\hat{H}_1 &= & -\frac12\tk\;\sum\limits_n \bigl\{e^{-i\hbar^*(n+1/2)}|n\rangle\langle n+1|
\nonumber \\
&& +\ e^{i\hbar^*(n-1/2)}|n\rangle\langle n-1|\bigr\},
\end{eqnarray}
and so
$$
\langle n|\tilt |0\rangle\;=\;\frac12(1-e^{i\hbar^*/2})\;\delta(n+1)\;+\;\frac12(1-e^{i\hbar^*/2})\;\delta(n-1)\;.
$$
Using this result, we have $\langle 0|\tilt|0\rangle=0$ and
\begin{eqnarray}
\langle 0|\tilt^2|0\rangle& =&  \sum\limits_{n\in\ZM}|\langle 0|\tilt|n\rangle|^2 \nonumber  \\
& =&  2\tk^2\bigr[1-\cos(\hbar^*/2)\bigr]
\sim\tk^2{\hbar^*}^2/4.
\end{eqnarray}

{\it Period of classical motion near separatrix} --
The classical period of motion along a tori close to the seperatrix was long known to be diverging as the initial condition approaches
the separatrix~\cite{Zas}. For self-completeness we give a detailed derivation here.  For convenience we consider the following Hamiltonian, which is canonically equivalent to the classical tHH of Eq.~(\ref{clthh}):
$$
H\;=\;\tk\;\bigl[\;\cos(q)\;-\;\cos(p)\;\bigr].
$$
The action variable at $H=E$ with $E<0$ is
\begin{gather}
I(E)\;=\;\frac1{2\pi}\oint p\;dq\nonumber\\
=\;\frac1{\pi}\int_{q_0}^{\pi}dq\;\arccos\bigr[\cos(q)-{E}{\tk}^{-1}\bigr]
\end{gather}
where $\cos(q_0)=1+E\tk^{-1}$ and so , as $E\to 0$,
$$
q_0\sim\;\sqrt{-2E\tk^{-1}}\;.
$$
Half-period at energy $E$ reads as
\begin{gather}
T(E)\;=\;\pi\biggl|\frac{dI(E)}{dE}\biggr|\nonumber\\
=\;\biggl|-\frac{dq_0}{dE}\arccos\bigl[-E\tk^{-1}+\cos(q_0)\bigr]\;+\nonumber\\
+\;\tk^{-1}\int_{q_0}^{\pi}dq\;\frac1{\sqrt{1-(-E\tk^{-1}+\cos(q))^2}}\biggr|\;.
\end{gather}
The first term vanishes, and the last integral splits as follows:
$$
\int^{\pi}_{q_0}dq\;\bigl(\ldots\bigr)\;=\;\biggl(\int_{q_0}^{\pi-q_0}\;+
\;\int_{\pi-q_0}^{\pi}\biggr)dq\;\bigl(\ldots\bigr)\;;
$$
Here the second integral is $\sim (\pi\tk)^{-1}$ as $E\to 0$, hence at leading order will be neglected. As to the first integral, expand at the first Taylor order in $E$ the integrand becomes
\begin{eqnarray}
\frac1{\sqrt{1-\bigl[\cos(q)-E\tk^{-1}\bigr]^2}} &\sim& \frac1{\sqrt{1-\cos^2(q)}} \nonumber \\
&& +\
\frac{E}{2\tk}\frac1{\sqrt{\bigl[1-\cos^2(q)\bigr]^3}}. \nonumber \\
\end{eqnarray}
Substituting this into the integral, the second term is found to yield a contribution which does not diverge in the limit $E\to 0$ and so will be neglected. The first term yields a contribution that rewrites as
\begin{gather}
\tk^{-1}\int_{q_0}^{\pi-q_0}dq\;\csc(q)\nonumber\\
=\;\tk^{-1}\ln\bigl[\tan(q)\bigr]\bigg|_{q_0/2}^{\pi/2-q_0/2}\nonumber\\
=\;-2\tk^{-1}\ln\biggl[\tan(\sqrt{-E(2\tk)^{-1}}\biggr]\nonumber\\
\sim\;\tk^{-1}\ln(-2\tk/E)\;,
\end{gather}
and so, putting all the above together, we have (at the leading order) the following
\begin{equation}
\label{cper}
T(E)\;\sim\;\tk^{-1}\ln(-2\tk/E)\;,
\end{equation}
which is exactly the Eq.~(7) we need.

\end{document}